\documentclass[a4paper,11pt]{article}
\pdfoutput=1 % if your are submitting a pdflatex (i.e. if you have
\usepackage{jheppub} % for details on the use of the package, please
\usepackage{graphicx}% Include figure files
\usepackage{dcolumn}% Align table columns on decimal point
\usepackage{bm}% bold math
\usepackage{hyperref}% add hypertext capabilities
\usepackage{graphicx,slashed,hyperref,color,subfig}
\usepackage{amssymb}
\usepackage[normalem]{ulem}
\usepackage[utf8]{inputenc}
\hypersetup{
   bookmarks=true,         % show bookmarks bar?
   unicode=true,          % non-Latin characters in Acrobat�s bookmarks
   pdftoolbar=true,        % show Acrobat�s toolbar?
   pdfmenubar=true,        % show Acrobat�s menu?
   pdffitwindow=false,     % window fit to page when opened
   pdfstartview={FitH},    % fits the width of the page to the window
   pdftitle={My title},    % title
   pdfauthor={Author},     % author
   pdfsubject={Subject},   % subject of the document
   pdfcreator={Creator},   % creator of the document
   pdfproducer={Producer}, % producer of the document
   pdfkeywords={keyword1} {key2} {key3}, % list of keywords
   pdfnewwindow=true,      % links in new PDF window
   colorlinks=true,       % false: boxed links; true: colored links
   linkcolor=blue,          % color of internal links (change box color with linkbordercolor)
   citecolor=magenta,        % color of links to bibliography
   filecolor=magenta,      % color of file links
   urlcolor=cyan           % color of external links
}
\usepackage{comment}
\def\lsim{\mathrel{\rlap{\lower4pt\hbox{\hskip1pt$\sim$}}
    \raise1pt\hbox{$<$}}}         %less than or approx. symbol
\def\gsim{\mathrel{\rlap{\lower4pt\hbox{\hskip1pt$\sim$}}
    \raise1pt\hbox{$>$}}}         %greater than or approx. symbol

\newcommand{\newc}{\newcommand}
\newc{\renewc}{\renewcommand}
\newc{\scp}{\textcolor{black}}
\newc{\sml}{\textcolor{black}}
\newc{\dyc}{\textcolor{black}}
\preprint{KIAS-P21052}

\title{Festina-Lente Bound on Higgs Vacuum Structure and Inflation}% Force line breaks with \\

\author[a]{Sung Mook Lee,}
\author[a]{Dhong Yeon Cheong,}
\author[a]{Sang Chul Hyun,}
\author[a, b]{Seong Chan Park,}
\author[c]{and Min-Seok Seo} 

\note{Correspondence: SCP \& M-SS}

\affiliation[a]{Department of Physics \& IPAP \& Lab for Dark Universe, Yonsei University, Seoul 03722, Korea}
\affiliation[b]{Korea Institute for Advanced Study, Seoul 02455, Republic of Korea}
\affiliation[c]{Department of Physics Education, Korea National University of Education,
Cheongju 28173, Republic of Korea}

\emailAdd{sungmook.lee@yonsei.ac.kr}
\emailAdd{dhongyeon@yonsei.ac.kr}
\emailAdd{bsg04103@yonsei.ac.kr}
\emailAdd{sc.park@yonsei.ac.kr}
\emailAdd{minseokseo57@gmail.com}

\abstract{
 The recently suggested Festina-Lente (FL) bound provides a lower bound on the masses of ${\rm U(1)}$ charged particles in terms of the positive vacuum energy. Since the charged particle masses in the Standard Model (SM) are generated by the Higgs mechanism, the FL bound provides a testbed of consistent Higgs potentials in the current dark energy-dominated universe as well as during inflation. We study the implications of the FL bound on the UV behavior of the Higgs potential for a miniscule vacuum energy, as in the current universe. We also present values of the Hubble parameter and the Higgs vacuum expectation value allowed by the FL bound during inflation, which implies that the Higgs cannot stay at the electroweak scale during this epoch.}

\begin{document} 
\maketitle
\flushbottom

\section{Introduction}

Can our understanding of the universe based on low energy dynamics be completed up to quantum gravity? This is one of main themes of the swampland program, which provides conjectured constraints on the low energy effective field theories (EFT) to be consistent with the UV completion of quantum gravity \cite{Vafa:2005ui}. (For recent reviews, see Refs.~\cite{Brennan:2017rbf, Palti:2019pca, Grana:2021zvf}. Also see Refs.~\cite{Park:2018fuj, Cheong:2018udx, Seo:2018abc} for specific inflationary models.) While many conjectures are based on string compactification, some of them are motivated based on a more generic quantum gravity context, including black hole (BH) physics.
For instance, although string theory claims that de Sitter (dS) space is unstable \cite{Obied:2018sgi} (see also Refs.~\cite{Andriot:2018mav, Garg:2018reu, Ooguri:2018wrx}), the universe may stay in quasi-dS for a sufficiently long enough time \cite{Seo:2019wsh, Cai:2019dzj}.
 Then (quasi-)dS can be approximated as a stable background and the dS-BH solutions as well as their thermal behavior can be used to set bounds on the low energy parameters (see, for example, Ref.~\cite{Cohen:1998zx} and also Ref.~\cite{Seo:2021bpb} for a discussion concerning dS instability).

If a dS-BH is super-extremal, i.e., the BH horizon is not generated inside the cosmological horizon, the BH singularity is naked. This has been widely considered to be forbidden as claimed by the cosmic censorship hypothesis \cite{Penrose:1969pc}, motivated by the predictability from the initial data as well as the observational consistency so far. In order to avoid the naked singularity, we can impose that the charged Nariai BH in which the BH horizon coincides with the cosmological horizon must discharge without losing sub-extremality.\footnote{On the other hand, imposing sub-extremality for BH much smaller than dS horizon size provides Weak Gravity Conjecture \cite{Arkani-Hamed:2006emk}.}
From this, one obtains the Festina Lente (FL) bound \cite{Montero:2019ekk, Montero:2021otb} : the mass $ m $ for {\it every} state of charge $q$ under $ {\rm U(1)} $ gauge invariance with coupling $ g $ satisfies 
\begin{align}
	\frac{m^{4}}{8 \pi \alpha q^2} \geq V \geq 0
\end{align}
where $ \alpha \equiv \frac{g^{2}}{4\pi} $ and $ V = 3 M_{P}^2 H^2$ ($ H $ : Hubble parameter) is the energy density governing Hubble  expansion. One direct implication is that the massless charged state is forbidden for nonzero $ V $. This bound also forbids unwanted charged black hole remnants.

In the Standard Model (SM) of particle physics, ${\rm U(1)}$ gauge invariance corresponds to electromagnetism, which is generated through the spontaneous breaking of electroweak gauge invariance via the Higgs mechanism. Then every $ {\rm U(1)} $ charged particle mass is  proportional to the Higgs vacuum expectation value (vev). Moreover, the cosmic energy density of the current universe is dominated by  dark energy $ V = \Lambda_{\text{DE}} \sim 10^{-120}M_{P}^{4} $. These two are consistent with the FL bound as the masses of the charged particles in the SM are all many orders larger than the Hubble scale of the current universe $ H_{0} \sim \mathcal{O}(10^{-60}M_{P})$ \cite{Montero:2021otb}.
  
  On the other hand, under the desert scenario which assumes the absence of any new physics up to the grand unification or the Planck scale, the self quartic coupling of the Higgs may vanish at UV and such near criticality of the Higgs potential raises the Higgs vacuum stability issue \cite{Sher:1988mj,Degrassi:2012ry,Hook:2014uia,Markkanen:2018pdo}. Also, depending sensitively on low energy SM parameters, the Higgs potential may develop another (meta-)stable vacuum at UV, at which the masses of the ${\rm U(1)}$ charged particles are enhanced.

  In this case, the FL bound is satisfied only if the vacuum energy density contribution from the Higgs potential is tuned such that even if it dominates the dark energy, its value is still tiny.
  The UV Higgs vev is fixed in accordance with this requirement. In this work, we investigate such constraints more concretely by considering possible UV behaviors of the Higgs potential \cite{Hook:2014uia,Espinosa:2015qea,Markkanen:2018pdo}.

 Moreover, it is plausible to postulate the inflationary epoch at the early stage of the  universe, where the universe was in a quasi-dS phase.
  This not only resolves the horizon and flatness problems \cite{Guth:1980zm, Linde:1981mu, Albrecht:1982wi}, but also explains how quantum fluctuations become the seed for large scale structures \cite{Mukhanov:1981xt, Mukhanov:1990me}.
Intriguingly, whereas we typically assume the reheating temperature $ T_{\text{reh}} $ after the inflation to be larger than the Big Bang nucleosynthesis (BBN) scale of $\mathcal{O}(10~\text{MeV})$, the vacuum energy in this case is too large to satisfy the FL bound if the electron mass remains the same or the Higgs vev is given by the electroweak scale.
This implies that during inflation, the Higgs potential should be stabilized at the UV vev, so that the electron mass (the lightest $ {\rm U(1)} $ charged particle in SM) becomes heavy to satisfy the FL bound. Regarding this, we discuss the restriction of the FL bound on the UV Higgs vev and the vacuum energy density during inflation.

The paper is organized as follows.
In Sec.~\ref{Section:Higgs Vacua}, we discuss the applicability of the FL bound for various possible UV behaviors of the Higgs and obtain the FL bound constraints on the structure of the Higgs potential.
In Sec.~\ref{Section:FL Bound and Inflation}, we present the FL bound constraints on the Higgs vev and the vacuum energy during the inflation, together with relevant cosmological observables.
We conclude in Sec.~\ref{Section:Conclusion}.

\section{Structure of Higgs Vacua}
\label{Section:Higgs Vacua}

\begin{figure}[t]
	\begin{center}
		\includegraphics[width= 0.7 \textwidth]{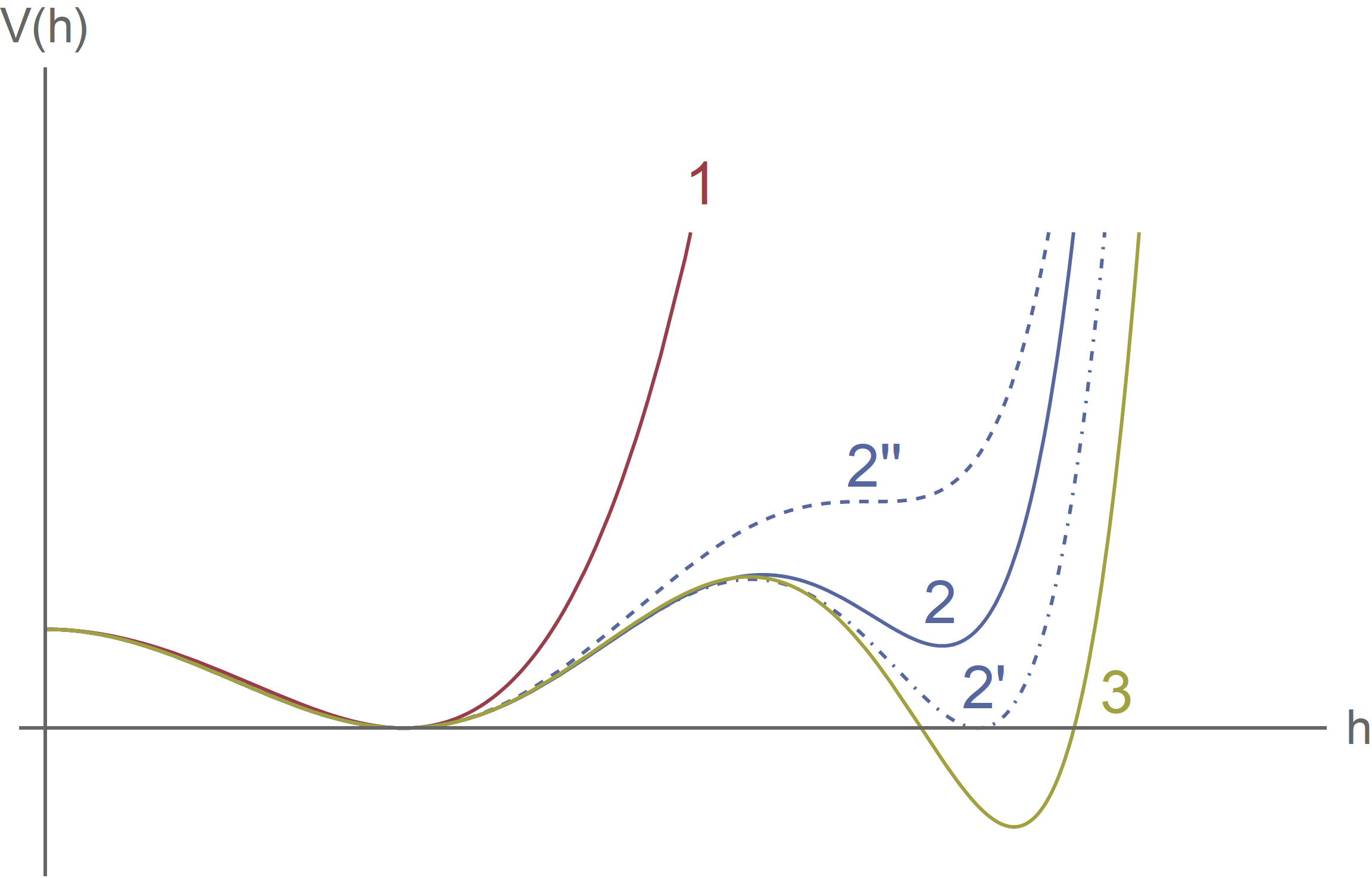}
		\caption{Schematic shape of the Higgs potential. \label{Figure:schematic}}
	\end{center}
\end{figure}

 The shape of the Higgs potential at UV ($ h \gg v_{\rm EW} $) that is consistent with low energy SM parameter values is not precisely determined, yet. The allowed shapes are schematically depicted in Fig.~\ref{Figure:schematic}.
 Even in the absence of new physics, the renormalization group (RG) running of the Higgs self quartic coupling $\lambda(\mu)$ depends sensitively on the SM parameters at electroweak (EW) scale, including ${\cal O}(1)$ top Yukawa coupling, i.e., the top quark mass.\footnote{In our definition, $ \lambda $ also includes the contribution from the correction to the 1PI effective action.}
 In addition, non-renormalizable operators that are irrelevant at the EW scale may become important at the UV scale as well. They allow various possible UV behaviors of the Higgs potential \cite{Degrassi:2012ry,Hook:2019zxa}.
  
Meanwhile, the FL bound comes from a gravitational argument, thus it is reasonable to assume that the FL bound can be applied to cases where the Higgs dominates the vacuum energy and the Higgs remains at some UV value for a sufficient amount of time. In this section, we discuss the applicability of the FL bound in these cases for each possible UV structure of the Higgs potential.
  
\subsection{Higgs Potential}

Imposing $ \mathbb{Z}_{2} $ symmetry, the UV Higgs potential we consider includes dark energy (DE), and higher order terms suppressed by cut-off scale $\Lambda$ as
 \begin{align}
	V_{\text{eff}}(h) & = \Lambda_{\text{DE}} +  \frac{\lambda(h)}{4} h^{4} + \frac{c_{6}}{\Lambda^{2}} h^{6} + \frac{c_{8}}{\Lambda^{4}} h^{8} + \cdots.
	\label{Eq:effectivepotential}
\end{align}
The constant DE $ \Lambda_{\text{DE}} \sim 10^{-120} M_{P}^{4} $ will be neglected in the succeeding discussion. The effective potential can be comprehensively written by defining the effective quartic coupling as
\begin{align}
	\lambda_{\text{eff}}(h) \equiv \frac{ 4 V_{\text{eff}} }{h^{4}}. 
\end{align}
Regarding the Higgs quartic coupling $\lambda(h)$ in Eq.~\eqref{Eq:effectivepotential}, we take the renormalization scale $\mu$ to be $h$ following Refs.~\cite{Degrassi:2012ry,Hamada:2014iga, Hamada:2014wna}. In the vicinity of some specific value $ h = h_{*} $, which we usually take $ h_{*}=h_{0} $ at which $ \lambda(h_{0}) = 0, $ %in the following discussion,
$\lambda(h) $ is expanded as 
\begin{align}
	\lambda(h) = \lambda (h_{*}) - \frac{b_{1}}{(4\pi)^{2}} \ln \frac{h}{h_{*}} + \frac{b_{2}}{(4\pi)^{4}} \ln^{2} \frac{h}{h_{*}} + \cdots.
\end{align}
If we consider the pure SM sector, $ b_{1} = -(4\pi)^{2} \beta_{\lambda}(h_{*}) \simeq 1/4 $ where $\beta_{\lambda} \equiv d \lambda/d (\ln \mu) $ works quite well for our purposes. In the non-renormalizable terms $ c_{2n} (n \geq 3) $ are unknown $\mathcal{O}(1)$ coefficients. These higher order terms are also required to impose the stability of the Higgs potential when $ \lambda(h) $ runs to negative values at high scales. In this work, to be explicit, we keep up to the dimension six operator setting $c_6=1$ and, $c_8 = c_{10} =\cdots =0$ letting the cutoff scale $\Lambda$ a free parameter.\footnote{When $c_{2n}$ with $ n \geq 4 $ are neglected, $c_6$ must be positive for the stability of the Higgs potential.}

Here, we schematically categorize the shapes of $V_{\rm eff}$ in three distinctive cases:
\begin{itemize}
	\item {\bf Case 1:} $ V_{\rm eff} $ has a unique EW vacuum at $ v_{\rm EW} $, hence the potential is monotonically increasing beyond $v_{\rm EW}$. As we will show, the FL bound cannot be applied for $ h > v_{\text{EW}} $ unless the potential is close to Case $ 2^{\prime\prime} $.
	
	\item {\bf Case 2:} $ V_{\rm eff} $ is positive for all $h$ and has another local minimum at UV, $ v_{\text{UV}} $. There are two interesting limits for this case.

\begin{itemize}

	\item Case $ 2^{\prime} $ :
$ V_{\rm eff} $ possesses a (almost) degenerate Higgs vacua, i.e.,
\begin{align}
	V_{\text{eff}} (v_{\text{UV}}) \simeq 0,&&	V_{\text{eff}}^{\prime} (v_{\text{UV}}) = 0,
\end{align}
and the same relations hold for $ v_{\rm EW} $.

	\item Case $ 2^{\prime\prime} $ : $ V_{\rm eff} $ has an inflection point at which
	\begin{align}
		V_{\text{eff}}^{\prime} (v_{\text{inf}}) = 0,&&	V^{\prime\prime} (v_{\text{inf}}) = 0.
	\end{align}

\end{itemize}
	
	\item {\bf Case 3:} $ V_{\rm eff} $ has a UV local minimum $ v_{\text{UV}} $, but becomes negative around $ v_{\text{UV}} $. In this case, the FL bound is not applicable because the universe would be AdS rather than dS unless  another source of vacuum energy is added. 
	We will visit this case in Sec. \ref{Section:FL Bound and Inflation}.
	 
\end{itemize}
 Since we restrict our discussion to $ \Lambda_{\rm DE} \simeq 0 $ in this section, we consider the FL implications on Case 1 and Case 2.

\subsection{Implications of FL bound}

 For the FL bound to be applied, by construction, the background of the universe  should be close to dS for an extended period of time compared to the BH lifetime.\footnote{More precisely, the background geometry after the charged BH production is required to be deformed  close to that of the Nariai BH, dS$_2\times$S$^2$ \cite{Montero:2019ekk, Montero:2021otb}, which undoubtedly includes the nearly constant cosmological horizon case. }
 This condition can be comprehensively written in terms of the potential slow-roll parameter defined by
 \begin{align}
 &\epsilon_{V} \equiv \frac{M_{P}^{2}}{2} \left( \frac{V_{\text{eff}}^{\prime}(h)}{V_{\text{eff}}(h)} \right)^{2}
= \frac{ 8 M_{P}^{2}}{h^2} \Big(1 + \frac{\beta_{\lambda,\rm eff}}{4 \lambda_{\rm eff}} \Big)^{2} 
\label{Eq:epsilon}
\end{align}
where $\beta_{\lambda,\rm eff} \equiv d \lambda_{\rm eff}/d (\ln \mu) $. With $g V^{1/2}$ being the electric field for the Nariai BH, we should have
\begin{align}
 \epsilon_{V}  \ll  e^{-\frac{m^2}{g q \sqrt{V}}} < 1.
\end{align}

\subsubsection{Case 1: Single Vacuum at $h=v_{\rm EW}$}

In Case 1, the Higgs potential has one local minimum at $ v_{\rm EW} $, then monotonically increases. The EW vacuum is consistent with the FL bound provided the dark energy is as small as $ \Lambda_{\text{DE}} \lesssim m_e^4/(8\pi\alpha) \sim 10^{-88} M_P^4$ since the electron is the lightest ${\rm U(1)}$ charged particle in the SM. This bound is consistent with the current universe  since $ \Lambda_{\text{DE}} \sim 10^{-120} M_{P}^{4} $ as  measured from Planck ~\cite{Planck:2018jri} satisfies the bound, which is noticed earlier in Ref.~\cite{Montero:2021otb}. 
The validity during inflation will be discussed in detail in Sec.~\ref{Section:FL Bound and Inflation}.

\subsubsection{Case 2: Two Vacua at $h=v_{\rm EW}$ and $v_{\rm UV}>v_{\rm EW}$}

We now turn our focus to Cases $2$, $2^\prime$, and $2^{\prime \prime }$, where the Higgs potential has a (meta-)stable UV vacuum $v_{\rm UV}$ satisfying $V^\prime(v_{\text{UV}}) = 0$ (hence $ \epsilon_{V} = 0 $) and $V(v_{\text{UV}}) \geq 0$.
Imposing the FL bound on the UV Higgs vacuum we obtain the  upper bound on the effective quartic coupling $ \lambda_{\text{eff}}(v_{\text{UV}})
> 0 $ as
\begin{align}
	\min_{i \in \text{SM}} \frac{m_{i}^{4}}{8\pi \alpha_{i}} = \frac{y_{e}^{4} v_{\text{UV}}^{4}/4}{8 \pi \alpha_{\text{EM}}} \geq  \frac{\lambda_{\text{eff}}(v_{\text{UV}})}{4} v_{\text{UV}}^{4} \\ \Rightarrow \quad \lambda_{\text{eff}}(v_{\text{UV}})  \leq \frac{y_{e}^{4}}{8 \pi \alpha_{\text{EM}}} \simeq \mathcal{O}\left( 10^{-22} \right).
\end{align}
Here the minimum of $m_i^4/\alpha_i$ is given by the SM electron which obtains its mass through the Higgs mechanism,
\begin{align}
	m_e (h)  = \frac{y_e(h)}{\sqrt{2}} h.
\end{align}
We note that there are several theoretical arguments justifying this seemingly fine-tuned small $ \lambda_{\text{eff}}(v_{\text{UV}}) $, which is out of the scope of this work (see, for example, Ref.~\cite{Kawai:2021lam}).

\subsubsection{Case $ 2^{\prime} $ : (Nearly) Degenerate Vacua, $v_{\rm EW} \simeq v_{\rm UV}$}

The degenerate case ($ V_{\text{eff}} \simeq 0 $  at $h_{1} = v_{\text{EW}}$ and $h_{2} = v_{\text{UV}}$)  satisfies the FL bound in a trivial manner. When $V_{\text{eff}}(v_{\text{UV}})$ is slightly uplifted, it does not violate the FL bound  provided $V$ is smaller then $10^{-22} v_{\rm UV}^4$, where we expect that $v_{\rm UV}$ in this case can be approximated by the values in the degenerate case.
Given 
\begin{align}
	\lambda(h) = \lambda_{*} - \frac{b_{1}}{(4\pi)^{2}} \ln \left( \frac{h}{h_{*}} \right), &&
	V_{\text{eff}}(h) = \frac{\lambda(h)}{4}h^{4} + \frac{1}{\Lambda^{2}} h^{6},
\end{align}
we estimate the values of $ v_{\rm UV} $ and the cut-off scale $ \Lambda $ from the conditions $ V_{\text{eff}}(v_{\text{UV}}) = 0 = V_{\text{eff}}^{\prime}(v_{\text{UV}}) $ as
\begin{align}
	v_{\text{UV}} &=h_{*}  \exp{\left(\frac{16 \pi ^2 \text{$ \lambda_{*} $}}{b_{1}}+\frac12\right)} \\
	\Lambda_{\text{UV}} & = \frac{8 \sqrt{2} \pi}{\sqrt{b_{1}}}  h_{*} \exp{ \left(\frac{16 \pi^{2} \lambda_{*}
		}{b}+ \frac{1}{2} \right)} = \frac{8\sqrt{2} \pi}{\sqrt{b_{1}}} v_{\text{UV}} . 
\end{align}
Note that both $v_{\rm UV}$ and $\Lambda_{\text{UV}}$ depend on the combination $h_{*} \exp \left( 16 \pi^2\lambda_{*}/b_{1} \right)$ which is nothing more than $ h \exp \left( 16 \pi^2\lambda(h)/b_1 \right)$.
Thus they are independent on the choice of the fiducial value $\lambda_{*}$, or equivalently $ h_* $ for the RG running, as expected.
Our estimation fixes the value of $ \lambda(v_{\rm UV}) $, which is physically meaningful, to be
 \begin{align}
	\lambda(v_{\text{UV}}) = - \frac{b_{1}}{32\pi^{2}}.
\end{align}
Since $ \lambda (v_{\text{UV}}) $ is negative, it is cancelled with $ (v_{\text{UV}}/\Lambda_{\text{UV}})^2 $ to give $ V_{\text{eff}}(v_{\text{UV}}) = 0 $ or $ \lambda_{\text{eff}}(v_\text{UV})  = \lambda(v_\text{UV}) + (v_\text{UV} / \Lambda_{\text{UV}})^2 = 0 $.

\begin{figure}[t]
	\centering
	\includegraphics[width=0.7\textwidth]{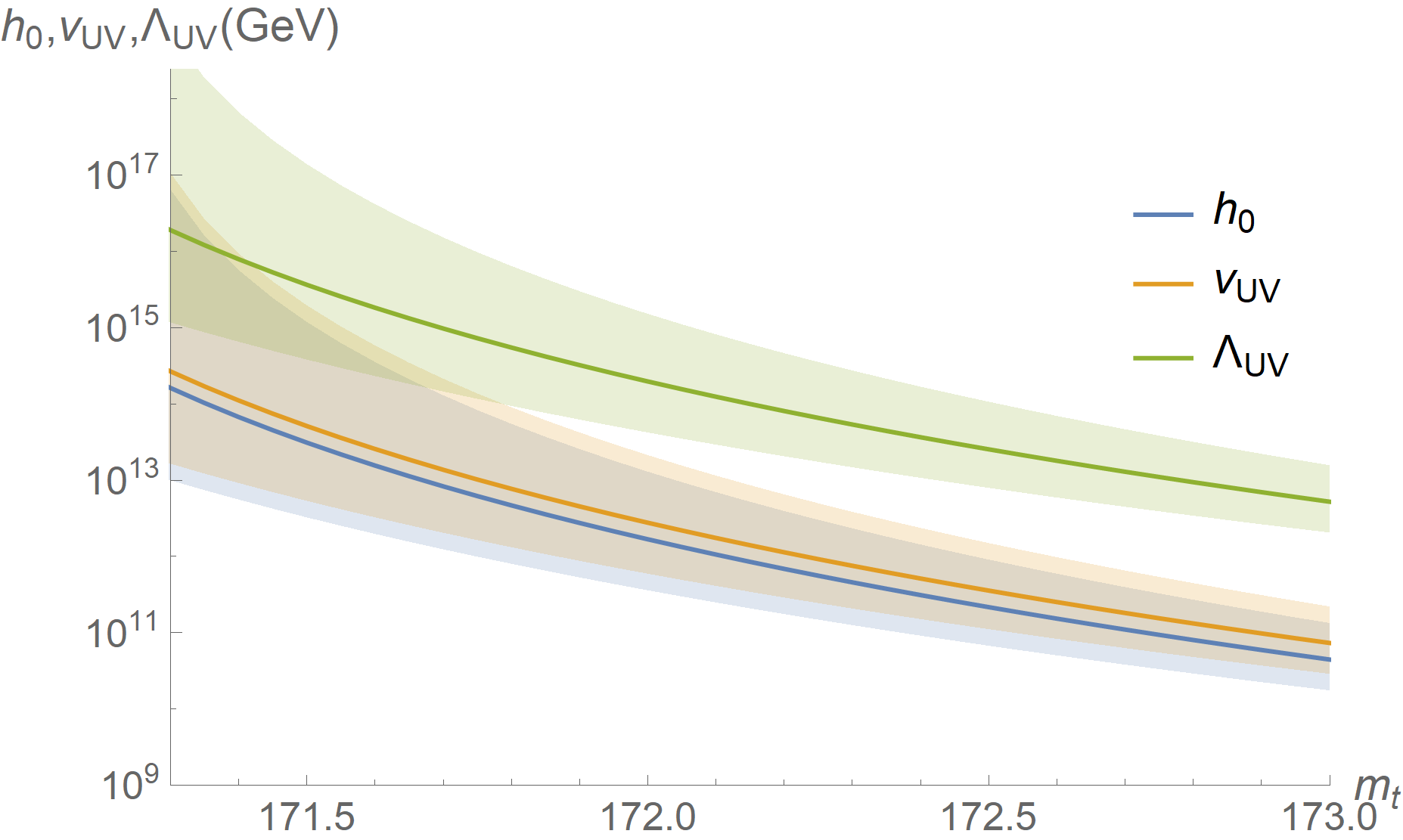}
	\caption{Scales for $ h_{0} $, $ v_{\text{UV}} $, $ \Lambda_{\text{UV}} $ depending on the top quark mass. Large additional uncertainty comes from strong coupling constant, which is depicted as shaded region at $ 1\sigma $ level with $ \alpha_{s} = 0.1179 \pm 0.0010 $ \cite{Zyla:2020zbs}. \label{Figure:running}}
\end{figure}

 One convenient choice of $ h_* $ is the value of the Higgs $ h_0 $ giving $\lambda (h_0) = 0 $, in terms of which we obtain
 \begin{align}
	v_{\text{UV}}  =  \sqrt{e} h_{0}, &&
	\Lambda_{\text{UV}}  =  \frac{8 \sqrt{2 e}\pi}{\sqrt{b_{1}}}h_0.
\end{align}
While the RG running of the SM parameters gives $ h_{0}  \gtrsim 10^{10}~\text{GeV}$, the explicit value of $h_0$ is sensitive to the top quark mass. Explicit values of ($ h_{0} $, $ v_{\text{UV}} $, $ \Lambda_{\text{UV}} $) depending on the EW top quark mass with uncertainty from strong coupling constant are shown in Fig.~\ref{Figure:running}.

\subsubsection{Case $2^{\prime\prime}$ : Inflection Point at $v_{\rm UV}$}

We now consider the Higgs potential shape close to Case $ 2^{\prime\prime} $ in which the inflection point exists.
 By setting  $\lambda_* = \lambda_0 =0$ we obtain from the conditions $ V_{\text{eff}}^{\prime}(v_{\rm inf})=0$ and  $ V_{\text{eff}}^{\prime\prime}(v_{\rm inf})=0$ that  $ v_{\rm inf} $ and the cutoff scale $\Lambda_{\text{inf}} $ are given by 
\begin{align}
	v_{\text{inf}} = e^{1/4} h_{0}, && \Lambda_{\text{inf}} = \frac{8\sqrt{3} \pi e^{1/4}}{\sqrt{b_{1}}} h_{0}.
\end{align}
Then the value of the potential height at $ v_{\rm inf} $ is,
\begin{align}
	V_{\text{eff}}(v_{\text{inf}}) = \frac{b_{1}e}{768 \pi^{2}} h_{0}^{4}.
\end{align} 
Since $ \lambda_{\text{eff}}(v_{\rm inf}) = \frac{b_{1}}{192\pi^{2}} \simeq \mathcal{O}(10^{-4}) \gg 10^{-22} $,  we conclude that the vacuum energy at the inflection point is too large to satisfy the FL bound.
 We also note that the inflection point is not consistent with the dS swampland conjecture, the refined version of which requires $\eta_V(v_{\rm inf})<-{\cal O}(1)$ even if $\epsilon_V(v_{\rm inf})\ll 1$ \cite{Andriot:2018mav, Garg:2018reu, Ooguri:2018wrx}.

{We also note that} the stability of the vacuum in Case 2 discussed in this section is purely classical. However, especially for setups near the inflection point, {tunneling via a Coleman-de Luccia or Hawking-Moss instanton} may induce the UV vacuum to decay faster than the charged BHs, such that the FL bound may not be applicable. This case includes the inflection point itself. Therefore, even with a large $ \lambda_{\text{eff}} $, we cannot rule out Case $ 2^{\prime\prime} $ by the FL bound. See Appendix.~\ref{Appendix:Hawking Moss Instanton near Inflection Point} for details.

\section{FL Bound and Inflation}
\label{Section:FL Bound and Inflation}

\begin{figure}[t]
	\centering
	\includegraphics[width=0.7\textwidth]{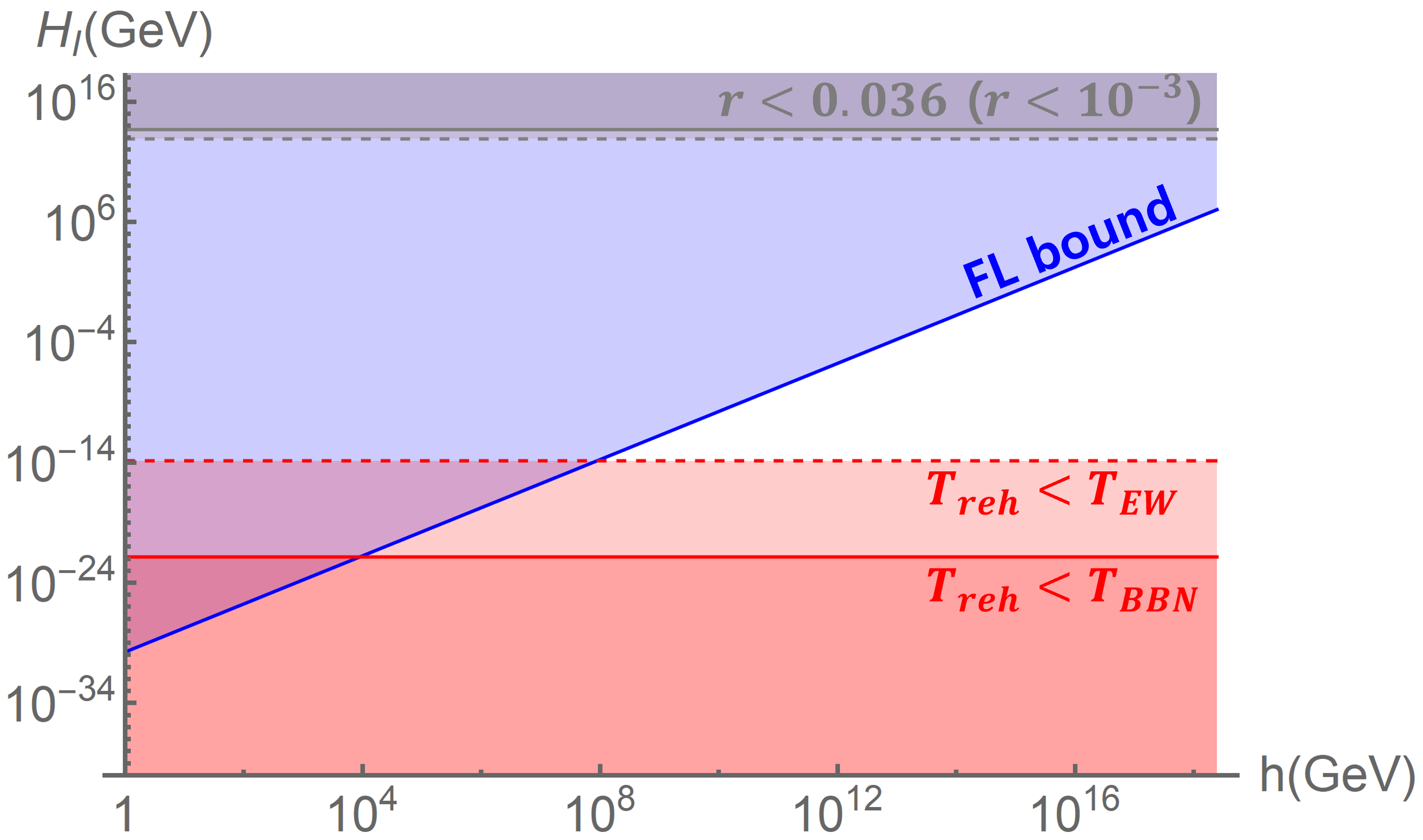}
	\caption{Bound on inflation scale from FL bound depending on the location of Higgs vev. Only white colored region is allowed. \label{Figure:bound}}
\end{figure}

In this section, we consider additional scalar fields $ \varphi_{i} $, $i=1,2,\cdots$, which, in addition to the Higgs field, contribute to the Hubble expansion especially during inflationary epoch:
 \begin{align}
	\sum_{i} U_{i}(\varphi_{i}) + V(h) =  3M_{P}^{2} H_{I}^{2},
\end{align}
where $ H_{I} $ is the (nearly constant) Hubble parameter. Then the FL bound is found taking the electron mass given by the Higgs mechanism with vacuum expectation value $ h $, as
 \begin{align}
	\frac{y_{e}^{4} h^{4}/4}{8 \pi \alpha_{\text{EM}}} \geq 3 M_{P}^{2} H_{I}^{2}
\end{align}
or
\begin{align}
	h \geq & \left(\frac{96 \pi \alpha_{\text{EM}}}{y_{e}^{4}}\right)^{1/4} \sqrt{M_{P} H_{I}}.
	\label{Eq:vUVbound}
\end{align}
The region excluded by the FL bound is colored in blue in Fig.~\ref{Figure:bound}.
For $ h < M_{P} $, using  $ y_{e}(M_{P}) \simeq 2.7 \times 10^{-6} $ and $ \alpha_{\text{EM}}(M_{P}) \simeq 9.5 \times 10^{-3} $,  we obtain a strong bound on the Hubble parameter during inflation
\begin{align}
	H_{I} \lesssim 10^{7} ~\text{GeV}, \label{Eq:BoundonHubble}
\end{align}
which is much more stringent than the conventionally known bound $H_I < 10^{13-14}~\text{GeV}$ from the recent Planck+BICEP/Keck 2018 observations \cite{BICEPKeck:2021gln}.

If the Higgs stays at the EW vacuum  during the inflation ($h =v_{\rm EW}$), the FL bound is satisfied provided $ \sum_{i} U_{i}(\varphi_{i}) + V(h) \lesssim 10^{-88} M_P^4 $, or equivalently, $ H_I \lesssim  10^{-44} M_P \simeq 10^{-26}~\text{GeV}$. 
However, this is in tension with the typical inflation scenario, which requires the reheating temperature $ T_{\rm reh} $ to be larger than the BBN scale.
To see this, recall that the instantaneous reheating temperature is determined under the situation where the inflation energy density is immediately converted into radiation, $ \rho_{\text{inf}} \simeq \rho_{\text{rad}}$, thus
\begin{align}
	3M_{P}^{2} H_{I}^{2} \simeq \frac{\pi^{2}}{30} g_{\text{reh}} T_{\text{reh,inst}}^{4},
\end{align}
where $ g_{\text{reh}} $ is the effective degree of freedom during reheating. In general, the actual reheating temperature is upper bounded as $ T_{\text{reh}} \leq T_{\text{reh,inst}} $. The FL bound implies a lower bound on the vacuum expectation value during inflation in terms of the reheating temperature
\begin{align}
	h \geq \left(\frac{16 \pi^{3} g_{\text{reh}} \alpha_{\text{EM}}}{15 y_{e}^{4}}\right)^{1/4} T_{\text{reh}}  \simeq  10^{6} \cdot  T_{\text{reh}}.
\end{align}
Requesting the reheating temperature to be larger than the BBN scale $T_{\text{reh}} \geq T_{\text{BBN}} \sim 10~\text{MeV} $, we obtain 
\begin{align}
h \gtrsim 10^4~{\rm GeV}, && \text{(during inflation)}
\end{align} 
which is much larger than $v_{\rm EW} \simeq 246~\text{GeV}$. Therefore, under reasonably acceptable conditions, the Higgs does not seem to stay at the electroweak vacuum during inflation.\footnote{If the Hubble scale is larger than  the Higgs vev, the spontaneously broken EW gauge invariance is restored.
The FL bound imposes that the Higgs vev to be larger than $H_I$ during the inflation, which excludes this possibility.}
The region ruled out by  this requirement is colored in red in Fig.~\ref{Figure:bound}. We also depicted the case with the more stringent requirement $T_{\rm reh} > T_{\rm EW} \sim 100~\text{GeV}$ assumed.

After inflation, EW vacuum may be restored during the early thermal history of our universe, without any violation of the FL bound due to the departure from the (quasi-) dS phase.

We also comment on the supersymmetric extension of the SM and its breaking mechanism. If we consider the  gravity mediation \cite{Nilles:1983ge} of the SUSY breaking to the SM sector, the Higgs soft mass is typically in the Hubble scale, $H_I$. While the Higgs quartic coupling in the minimal supersymmetric extension is given by the gauge coupling, it can vanish when the Higgs is on the D-flat direction.  Then the quartic term is dominated by $\lambda \sim H_I^2/\Lambda^2$ such that the Higgs potential during inflation, schematically written in the form of
\begin{align}
	V(h) =	-\frac{c_2^2}{2}   H_I^2 h^2 + \frac14 \frac{H_I^2}{\Lambda^2} h^4, && c_{2} > 0
\end{align}   
is stabilized at $v_{\rm UV} = c_2 \Lambda$, which can be larger than Eq.~\eqref{Eq:vUVbound} thus satisfies the FL bound.
For gauge mediation \cite{Giudice:1998bp}, on the other hand, the soft mass $H_I M_P/M $ can be enhanced by the sub-Planckian messenger scale $M$, such that the Higgs vev satisfies Eq.~\eqref{Eq:vUVbound} for $ M < ( H_I M_P )^{1/2} $ even with a ${\cal O}(1)$ Higgs quartic coupling.

The bounds on $H_I$ and $h$ we obtained above is consequentially encoded in bounds of inflationary observables, which enables us to test the FL bound. The scalar(tensor) power spectrum of the quantum fluctuation during inflation is usually parametrized by the scalar(tensor) amplitude $ A_{s} (A_{t}) $ and spectral indices $ n_{s} (n_{t})$ with a pivot scale $ k_{*} $ as
\begin{align}
	\mathcal{P}_{\mathcal{S}} (k) = A_{s} \left( \frac{k}{k_{*}}\right)^{n_{s}-1}, &&
	\mathcal{P}_{\mathcal{T}} (k) = A_{t} \left( \frac{k}{k_{*}}\right)^{n_{t}},
\end{align}
and the tensor-to-scalar ratio is defined by $  r \equiv A_{t}/A_{s} $. The Planck+BICEP/Keck 2018 result provides $ A_{s} \simeq 2.1\times 10^{-9} $ and a bound on $ r < 0.036  $ with $ k_{*} = 0.05 ~\text{Mpc}^{-1}$ \cite{Planck:2018jri,BICEPKeck:2021gln}.

For the single field slow-roll inflation, the bound on the Hubble scale is directly converted to the bound on the tensor-to-scalar ratio using the relation
\begin{align}
	\frac{H_{I}^{2}}{M_{P}^{2}} \simeq \frac{\pi^{2}}{2} A_{s} r
\end{align}
as
\begin{align}
	r \lesssim 3 \times 10^{-15} \left( \frac{10^{-2}}{\alpha_{\text{EM}}}\right)  
	\left( \frac{2\cdot 10^{-9}}{A_{s}}\right) 
	\left( \frac{y_{e}}{3 \cdot 10^{-6}}\right)^{4} \left( \frac{h}{M_{P}}\right)^{4},
\end{align}
which is many orders smaller than the observability of current/proposed experiments \cite{BICEPKeck:2021gln, CMB-S4:2016ple, CMB-S4:2020lpa, LiteBIRD:2020khw}.

\section{Conclusion}
\label{Section:Conclusion}

In this work,  we analyze the implications of the recently proposed FL bound on the Higgs vacuum structure and the inflationary cosmology.
Among the possible structures the UV Higgs potential may have (as shown in Fig.~\ref{Figure:schematic}), the FL bound can be applied to Case 2, in which another positive local minimum appears at UV. The FL bound restricts $ V_{\text{eff}}(v_{\text{UV}}) $ to be almost vanishing such that the  UV vacuum is nearly degenerate with the EW vacuum. The sizes of  $ v_{\rm UV} $ and $ \Lambda_{\rm UV} $ are also constrained, the exact values of which sensitively depend on the top quark mass, as well as the strong coupling constant.

Meanwhile, unlike the current universe, the EW vacuum is not compatible with the vacuum energy when the universe was in  the  inflationary phase provided the reheating temperature is larger than the BBN scale. To satisfy the FL bound, the Higgs should have a vacuum at UV scale, satisfying $ h \gtrsim 10^{4} ~ \text{GeV} $ and the Hubble scale $H_I$ is required to be 
less than the order of $ 10^{7}~\text{GeV}$, implying minuscule tensor-to-scalar ratio $ r \lesssim 3 \times 10^{-15} $.

\acknowledgments

We are grateful to Misao Sasaki and Chang Sub Shin for discussions and valuable comments.
This work was supported by National Research Foundation grants funded by the Korean government (MSIT) (NRF-2019R1A2C1089334), (NRF-2021R1A4A2001897) (SCP), (MOE) (NRF-2020R1A6A3A13076216) (SML),  and (NRF-2021R1A4A5031460) (MS). The work of SML is supported by the Hyundai Motor Chung Mong-Koo Foundation Scholarship.

\appendix
\section{Hawking-Moss and Coleman-de Luccia Instanton near Inflection Point}
\label{Appendix:Hawking Moss Instanton near Inflection Point}

In this Appendix, we quantitatively estimate the UV vacuum transition rate to EW vacuum near the inflection point and find the conditions for the Higgs potential to be stable, even quantum mechanically.

For a given decay rate, we denote $ \Gamma \simeq A e^{-B} $. From FL, the Nariai BH decay rate is \cite{Montero:2019ekk,Montero:2021otb}
\begin{align}
	\Gamma_{\text{BH}} \sim \exp \left( - \frac{m^{2}}{gq \sqrt{V}}    \right) \sim \exp \left( - \frac{m^{2}}{g q M_{P} H}    \right).
\end{align}
To ensure the FL bound, stability of UV vacuum should be guaranteed, i.e. $ \Gamma_{\text{BH}} \gg \Gamma_{\text{dS}} $, hence $ B_{\text{BH}} \ll B_{\text{dS}}  $.

For the inflection point, we have
\begin{align}
	v_{\text{inf}} = e^{1/4} h_{0}, && \Lambda_{\text{inf}} = \frac{8\sqrt{3} \pi e^{1/4}}{\sqrt{b_{1}}} h_{0}= \frac{8\sqrt{3} \pi}{\sqrt{b_{1}}} v_{\text{inf}}.
\end{align}
with
\begin{align}
	V_{\text{inf}} 
	= \frac{b_{1}}{768 \pi^{2}} v_{\text{inf}}^{4}, && H = \frac{\sqrt{b_{1}} v_{\text{inf}}^{2}}{48\pi M_{P}}
\end{align}
and
\begin{align}
	\lambda_{\text{eff}} = \frac{4V_{\inf}}{v_{\inf}^{4}} = \frac{b_{1}}{192 \pi^{2}} \sim 10^{-4}.
\end{align}
Small deviation of $ \Lambda_{\text{inf}} $ as
\begin{align}
	\Lambda = \Lambda_{\inf} (1 + \delta_{\Lambda})
\end{align}
induces deviations for the field values at local maximum and local minimum
\begin{align}
	h_{\max} \simeq v_{\text{inf}} ( 1 - \delta_{v}), && h_{\min} \simeq v_{\text{inf}}(1+\delta_{v}) 
\end{align}
and $\Delta h \equiv h_{\text{max}} - h_{\text{min}}  \simeq 2 v_{\rm inf} \delta_v$ with $ \delta_{v} = \sqrt{\delta_{\Lambda}} $ at the lowest order. Putting this,
\begin{align}
	\delta_{\lambda} \equiv	\frac{\delta \lambda_{\text{eff}}}{\lambda_{\text{eff}}} = - 4 \delta_{v}, \,\, 
	\end{align}
and
\begin{align}
	\Delta V \simeq \frac{b_{1} v_{\inf}^{4}}{12\pi^{2}}\delta_{v}^{3}.
\end{align}

Near the inflection point, the relevant vacuum transition process is due to Coleman-de Luccia (CdL)
and/or Hawking-Moss (HM) instanton respectively with
\begin{align}
 B_{\rm CdL} \simeq \frac{2\pi^2 \sqrt{2\Delta V} (2\Delta h)}{ H^3}, &&
 B_{\text{HM}} \simeq \frac{8\pi^{2} \Delta V}{3 H^{4}}
\label{Eq:instanton}
\end{align}
\cite{HenryTye:2008xu, Hook:2014uia}. Since the ratio of the two quantities is given as
\begin{align}
\frac{B_{\rm CdL}}{B_{\rm HM}} \simeq \frac{3H \Delta h}{ \sqrt{2\Delta V}} \simeq \frac{\sqrt{3}}{4\sqrt{2}  \sqrt{\delta_v}} \left(\frac{v_{\rm inf}}{M_P}\right),
\end{align}
the decay of false vacuum is dominated by CdL or HM instanton depending on the numerical value of $\delta_v$. 
When $\delta_v \simeq  \delta_v^{\rm eq} =\frac{3}{32} \left(\frac{v_{\rm inf}}{M_P}\right)^2 $, $B_{\rm HM} \simeq B_{\rm CdL}$ thus both are equally important.\footnote{
{We also note that the HM solution contributes to the vacuum decay provided $ H > H_{\text{crit}} $ with
\begin{align}
	H_{\text{crit}} \equiv \sqrt{- \frac{V^{\prime\prime}(v_{\max})}{4} - \frac{\Delta V}{3M_{p}^{2}}} \simeq \frac{\sqrt{b_{1}} v_{\text{inf}}}{4\sqrt{2} \pi} \sqrt{\delta_{v}}, && \delta_{v} < \frac{1}{72} \left( \frac{v_{\text{inf}}}{M_{P}}\right)^{2} \equiv \delta_{v}^{\text{crit}}
\end{align}
\cite{Coleman:1985rnk,Brown:2007sd}, while for $ H < H_{\text{crit}} $ the CdL solution  contributes exclusively. Our estimation in Eq.~\eqref{Eq:instanton} is consistent to the fact that CdL and HM solutions usually coincide at the limit of $ H \rightarrow H_{\text{crit}} $ as $ \delta_{v}^{\text{eq}} \simeq \delta_{v}^{\text{crit}} $ \cite{Balek:2004sd}.}}

However, this result does not contain stochastic effects which are usually caught by Fokker-Planck (FP) approach. 
If $ H^{2} \lesssim \sqrt{\Delta V} $ corresponding to
\begin{align}
	\delta_{v} \gtrsim \frac{1}{48}  \left( \frac{b_{1}}{4\pi^{2}} \right)^{1/3}
	 \left(\frac{v_{\inf}}{M_{P}}\right)^{4/3} \simeq  4 \times 10^{-3} \left(\frac{v_{\inf}}{M_{P}}\right)^{4/3},
	 \label{Eq:HM}
\end{align}
only considering a single CdL or HM would be enough \cite{Hook:2014uia}. 
On the other hand, if $ H^{2} \gtrsim \sqrt{\Delta V} $, one has to consider FP process more carefully to determined the stability of the UV vacuum. For our purpose to see the smallness of the correction, it suffices to only consider the single CdL or HM  instanton process.

For this case, to guarantee the stability of UV vacuum, we should have
\begin{align}
B_{\text{BH}} = \frac{m^{2}}{gq M_{P} H} \ll \begin{cases}
B_{\rm CdL} \simeq \frac{2^{14} 3^2\sqrt{6} \pi^4}{b_1}  \left(\frac{M_P}{v_{\rm inf}}\right)^3 \delta_v^{5/2}, \\
B_{\text{HM}}\simeq \frac{2^{17}3^{2} \pi^{4}}{b_{1}} \left(\frac{M_{P}}{v_{\text{inf}}}\right)^{4} \delta_{v}^{3}
\end{cases} 
\end{align}
implying
\begin{align}
	\delta_{v} \gg
	\begin{cases}
	\left(\frac{ b_1 y_e^4 v_{\rm inf}^6}{2^{23} 3^3 \pi^7 \alpha_{\rm EM} M_P^6}\right)^{1/5} \simeq 3\times 10^{-7} \left(\frac{v_{\rm inf} }{M_P}\right)^{6/5} & \text{(CdL dominant)}\\
	\left( \frac{b_{1} y_{e}^{4} v_{\inf}^{8}}{2^{28}3^{2}\pi^{7} \alpha_{\text{EM}} M_{P}^{8}}\right)^{1/6}\simeq 2 \times 10^{-6} \left(\frac{v_{\text{inf}}}{M_{P}}\right)^{4/3} & \text{(HM dominant)} 
	\end{cases}		
\end{align}
As expected, dS vacuum transition rates are suppressed as $ \delta_{v} $ increase. Therefore, under the consideration of CdL and HM instanton, the UV vacuum is stable as long as Eq.~\eqref{Eq:HM} holds. This already shows that our results numerically do not change much even when we consider the quantum effects.

%======================================================================
%{\bf Acknowledgements:}

\bibliography{HiggsFL}
\bibliographystyle{apsrev4-1}
%\nocite{*}

\end{document}